\documentclass[aps,prb,twocolumn,superscriptaddress,nofootinbib,longbibliography,amsmath,amssymb]{revtex4-2}

\usepackage{graphicx}
\usepackage{bm}
\usepackage[colorlinks=true,linkcolor=blue,citecolor=blue,urlcolor=blue]{hyperref}

\newcommand{\Pzz}{P_{zz}}
\newcommand{\wX}{\omega_X}
\newcommand{\wH}{\omega_H}
\newcommand{\wQ}{\omega_Q}
\newcommand{\woX}{\omega_{1X}}
\newcommand{\woH}{\omega_{1H}}
\newcommand{\weff}{\omega_{\rm eff}}
\newcommand{\dd}{\mathrm{d}}
\newcommand{\degree}{^\circ}

\begin{document}

\title{Rotor-driven piecewise cross polarization of wide quadrupolar lines:\\
bath-contact pumping of vector and tensor order in dynamically polarized solids}

\author{D. Keller}
\author{M. F. Hossain}
\affiliation{University of Virginia, Charlottesville, USA}

\date{\today}

\begin{abstract}
In a dynamically polarized solid with a common nuclear spin temperature, spin-1 $^{14}$N in ammonia sits at $P_N\simeq0.14$ when the protons are at $0.90$, and a deuteron in a proton-rich matrix at $P_D\simeq0.29$. Hartmann--Hahn cross polarization (CP) from the protons could raise either toward $P_H$, but the first-order quadrupole interaction spreads the $^{14}$N line over $5$~MHz at $5$~T, far beyond any single contact bandwidth. We develop the theory of CP into one transition of a quadrupolar spin-1, with slow sample rotation carrying each crystallite through a fixed-frequency contact window. A bath contact is an idempotent map on the populations of the contacted transition, so repeated or continuous contacts accumulate rather than cancel, and the hit-once and parity constraints that force stepped rotation for pulsed inversions do not arise. In the small-reservoir limit the population ratio across the contacted transition becomes $(1+P_H)/(1-P_H)$, giving $\Pzz=+0.88$ in one contact and $\Pzz\to-1.8$ in two from $P_H=0.9$, beyond the unitary bound $[-0.85,+0.65]$. Under continuous rotation, a window at the Pake horn reaches every bond for an axis perpendicular to the field, while a window in the clean pedestal region, needed for a definite sign of tensor order, reaches a fraction $\sin2\theta_c=0.943$ at the magic angle $54.74\degree$. Steady-state order is set by the contact rate against rotation-driven cross-branch relaxation, negligible for $^{14}$N. In the cubic limit one contact places the spin-1 at the proton temperature ($\Pzz=+0.85$), but only a resolved quadrupole splitting permits negative tensor order. The limiting resource is the proton spin lock, which must far exceed the $\simeq28$~kHz local field for a contact short compared with $T_{1\rho}$, unmeasured at $1$~K. Protocols are given for $^{14}$N in NH$_3$ and deuterons in NH$_2$D.
\end{abstract}

\maketitle

\section{Introduction}

\begin{figure}
\includegraphics[width=\columnwidth]{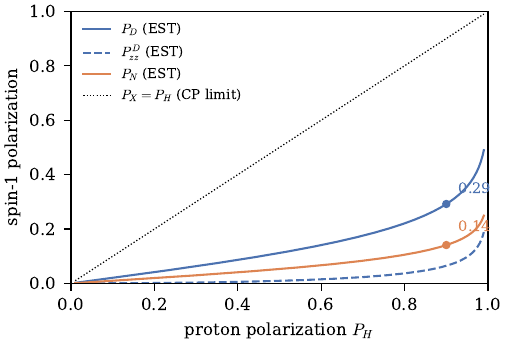}
\caption{Equal-spin-temperature polarization of spin-1 nuclei versus proton polarization, Eq.~\eqref{eq:est}: deuteron vector and tensor polarization and $^{14}$N vector polarization. The dotted line is the small-reservoir CP limit $P_X=P_H$. Markers: $P_D=0.29$ and $P_N=0.14$ at $P_H=0.90$.}
\label{fig:est}
\end{figure}

Dynamic nuclear polarization (DNP) in the solid state cools the nuclear Zeeman reservoirs of all species present to a common spin temperature through the electron dipolar reservoir \cite{borghini1968,abragamgoldman1978,goldman1970}. Equal spin temperature (EST) is well established experimentally: for $^{14}$N and protons in irradiated ammonia \cite{adeva1998,court1986,friedman2019}, and for protons and deuterons in partially deuterated alcohols \cite{deboer2015}. Its consequence for a low-$\gamma$ nucleus is unfavourable. Writing $a_X=\gamma_X\hbar B_0/k_BT_s$ for the Zeeman energy of species $X$ in units of the spin temperature, the spin-1 polarization is $P_X=2\sinh a_X/(1+2\cosh a_X)$ and $a_X/a_H=\gamma_X/\gamma_H$, so that at $P_H=0.90$ the $^{14}$N nuclei of ammonia sit at $P_N=0.14$ and deuterons in a proton-dominated matrix at $P_D=0.29$ (Fig.~\ref{fig:est}).

Cross polarization (CP) \cite{hartmann1962,lurie1964,pines1973} offers a way past this. At the Hartmann--Hahn match the contacted transition of $X$ equilibrates with the proton reservoir in the rotating frame, and in the small-reservoir limit its population ratio becomes the proton ratio $(1+P_H)/(1-P_H)$: at $P_H=0.9$ a ratio of $19$, against the EST ratio of $e^{a_X}$, which is $1.57$ for deuterons and $1.24$ for $^{14}$N. The obstacle is bandwidth. The first-order quadrupole interaction spreads the $^{14}$N line of ammonia, with $e^2qQ/h=3.47$~MHz \cite{lehrer1965}, over $5.2$~MHz at $5$~T, and the deuteron line over several hundred kHz, while a CP contact addresses a window of order $\woX$, a few tens of kHz. Broadband variants, adiabatic-sweep CP (BRAIN-CP) \cite{harris2012} and its relatives \cite{kupce1995}, extend the window to a few hundred kHz, and frequency stepping \cite{massiot1995} covers the rest at the cost of retuning the probe at every step. Nothing covers a $5$~MHz line in one contact.

This paper develops an alternative: keep the contact at one frequency and rotate the sample slowly so that every crystallite is carried through the window, in the manner of the slowly rotating, continuously saturated targets of Keller \emph{et al.} \cite{keller2020}. The companion paper \cite{companion} analysed sample rotation combined with coherent inversion pulses and found it severely constrained: inversion is a unitary involution, so each crystallite may be addressed exactly once, the stops must be discrete and packed as caps on the orientation sphere, the transits must be fast enough to be diabatic for dipolar pair crossings, and a parity theorem forbids any gain from leaving the rf on during rotation. All of this disappears for CP. A bath contact is an idempotent projection on the populations of the contacted transition, so repeated contacts accumulate, continuous rotation with continuous contact is allowed, the natural geometry is the union rather than the symmetric difference of caps, and the rotation-driven dipolar flip-flops that scramble an inversion pattern are harmless because every contact resets the matched bonds. The constraints that remain are of a different kind: the lifetime of the proton spin lock, the proton local field it must overcome, the reservoir heat capacities, and, for deuterons, cross-branch relaxation between contacts.

Section~\ref{sec:system} sets up the two-species spin system, the EST state and the chemistry of the candidate materials. Section~\ref{sec:cp} derives CP into one transition of a spin-1 with the single-transition operator formalism \cite{vega1978,vega1977,wokaun1977}, the rotating-frame thermodynamics \cite{redfield1955,slichter1961,lurie1964}, the bath-contact map and its idempotence, the heat-capacity correction and the adiabatic (rotor-driven) form of the contact. Section~\ref{sec:geometry} derives in closed form which bonds a fixed window reaches under continuous rotation about an axis at angle $\beta$, and the optimal $\beta$ for horn and pedestal windows. Section~\ref{sec:rates} treats the dwell time in the window, the contact train, the proton lock, and the steady state against relaxation. Section~\ref{sec:protocols} gives protocols for $^{14}$N in NH$_3$ and for deuteron tensor order in NH$_2$D, Sec.~\ref{sec:tests} the measurements that decide them, Sec.~\ref{sec:materials2} the cubic limit $\wQ\to0$ and the material requirements that follow from it, with ammonium salts as candidates, and Sec.~\ref{sec:conclusion} concludes.

\section{The two-species spin system}
\label{sec:system}

\subsection{Hamiltonian}

Let $X$ be a spin-1 nucleus ($^{14}$N or $^2$H) with an axially symmetric electric field gradient along the bond, and $H$ the protons. In a field $B_0\hat z$,
\begin{align}
\mathcal H=&-\wH\sum_j I_{zj}-\wX\sum_k S_{zk}+\sum_k\wQ c_k\,(3S_{zk}^2-2)\nonumber\\
&+\mathcal H_{HH}+\mathcal H_{HX}+\mathcal H_{XX},
\label{eq:H}
\end{align}
with $c_k=3\cos^2\theta_k-1$, $\theta_k$ the angle between bond $k$ and $B_0$, $\wQ=e^2qQ/8\hbar$, and the dipolar terms \cite{abragam1961}
\begin{align}
\mathcal H_{HH}&=\sum_{i<j}b_{ij}\bigl[I_{zi}I_{zj}-\tfrac14(I_i^+I_j^-+I_i^-I_j^+)\bigr],\nonumber\\
\mathcal H_{HX}&=\sum_{jk}d_{jk}\,I_{zj}S_{zk},\nonumber\\
\mathcal H_{XX}&=\sum_{k<l}b'_{kl}\bigl[S_{zk}S_{zl}-\tfrac14(S_k^+S_l^-+S_k^-S_l^+)\bigr],
\label{eq:dip}
\end{align}
where the heteronuclear coupling is truncated to its secular $I_zS_z$ part because $\wH\ne\wX$, and $d_{jk}=(\mu_0/4\pi)\gamma_H\gamma_X\hbar\,r_{jk}^{-3}(1-3\cos^2\vartheta_{jk})$. The single-quantum transitions of $X$ are at $\wX\mp3\wQ c_k$ and, in the reduced coordinate $R=(\omega-\wX)/3\wQ$, follow the Pake densities $g_\pm(R)=[2\sqrt{3(1\mp R)}]^{-1}$ on $[-2,1]$ and $[-1,2]$ \cite{pake1948,cohen1957}; the companion paper \cite{companion} gives the lineshape in detail. Two regions of the line matter here: the \emph{horn} $|R|=1$, $\theta=90\degree$, where the powder density diverges and both branches are present, and the \emph{clean pedestals} $1<|R|\le2$, $\theta<\theta_c=\arccos\sqrt{2/3}=35.26\degree$, where only one branch is present.

\subsection{The EST state}
\label{sec:est}

Under DNP all nuclear Zeeman reservoirs reach a common inverse spin temperature $\beta_s$ when their thermal contact with the electron dipolar reservoir is good \cite{borghini1968,abragamgoldman1978,deboer2015}, so that each species has populations $p_m\propto e^{m a_X}$ with $a_X=\beta_s\hbar\wX$. For spin $\tfrac12$ protons $P_H=\tanh(a_H/2)$, and for spin-1
\begin{align}
P_X&=\frac{2\sinh a_X}{1+2\cosh a_X},\qquad
\Pzz^X=\frac{2\cosh a_X-2}{1+2\cosh a_X},\nonumber\\
a_X&=\frac{\gamma_X}{\gamma_H}\,2\operatorname{artanh}P_H .
\label{eq:est}
\end{align}
Table~\ref{tab:est} gives the values for the two nuclei of interest. The spin temperature actually reached is set by the balance between the cooling of the electron dipolar reservoir and the leakage of all nuclear reservoirs to the lattice, and the leakage is dominated by the majority species: ND$_3$ reaches $P_D\simeq0.5$, which corresponds to a colder spin temperature ($a_H=5.5$, a proton polarization of $0.99$) than NH$_3$ at $P_H=0.9$ ($a_H=2.9$). In a mixed material such as NH$_2$D the protons dominate the heat capacity and the leakage, so the NH$_3$-like temperature is the expectation and $P_D\simeq0.29$ follows; this has been verified for partially deuterated propanediol \cite{deboer2015} but not for ammonia, where no polarized-target data on NH$_2$D or NHD$_2$ exist.

\begin{table}
\caption{EST polarizations, Eq.~\eqref{eq:est}, and the populations $(p_+,p_0,p_-)$ of the spin-1 species at $P_H=0.90$ ($a_H=2.944$).}
\label{tab:est}
\begin{ruledtabular}
\begin{tabular}{lcccc}
$X$ & $\gamma_X/\gamma_H$ & $a_X$ & $P_X$, $\Pzz^X$ & $(p_+,p_0,p_-)$ \\
\hline
$^2$H & $0.1535$ & $0.452$ & $0.292$, $0.065$ & $(0.490,0.312,0.198)$ \\
$^{14}$N & $0.0723$ & $0.213$ & $0.141$, $0.015$ & $(0.402,0.328,0.270)$ \\
\end{tabular}
\end{ruledtabular}
\end{table}

\subsection{Materials}
\label{sec:materials}

The candidate materials are the ammonia isotopologues NH$_3$, NH$_2$D, NHD$_2$ and ND$_3$, which polarize well after irradiation \cite{meyer2004,crabb1997,goertz2002}. Solid ammonia is cubic (P$2_13$) with hydrogen-bonded molecules, N--H $=1.01$~\AA, H--N--H $=107\degree$ and a nearest N--N distance of $3.35$~\AA; the $^{14}$N site has $C_{3v}$ symmetry so $\eta=0$, with $e^2qQ/h=3.47$~MHz in both NH$_3$ and ND$_3$ \cite{lehrer1965}. The deuteron coupling in ND$_3$ is of order $200$~kHz (N--D, approximate). Table~\ref{tab:couplings} lists the dipolar couplings that enter: the geminal H--H coupling of $28$~kHz sets the proton local field that a spin lock must overcome; the H--N and H--D couplings of $8.4$ and $4.3$~kHz set the CP transfer rate; and the X--X couplings set the rotation-driven relaxation of Sec.~\ref{sec:rates}, which is negligible for $^{14}$N ($17$~Hz between neighbouring molecules) and significant for deuterons only when geminal D--D pairs are present. The isotopologue therefore selects the physics: NH$_3$ for $^{14}$N; NH$_2$D for deuterons, which has no geminal D--D pair and whose nearest D--D distance of $\simeq3$~\AA{} gives a coupling of $\simeq0.1$~kHz; NHD$_2$, whose geminal D--D coupling of $0.67$~kHz exceeds that of C--D in butanol, is the worst choice for deuteron tensor work.

\begin{table*}
\caption{Dipolar couplings $|b|/2\pi$ in ammonia with the angular factor set to unity, $(\mu_0/4\pi)\gamma_1\gamma_2\hbar/2\pi r^3$.}
\label{tab:couplings}
\begin{ruledtabular}
\begin{tabular}{lccc}
Pair & $r$ (\AA) & $|b|/2\pi$ (kHz) & Role \\
\hline
H--H geminal & $1.62$ & $28.3$ & proton local field, lock requirement \\
H--N & $1.01$ & $8.4$ & CP rate, $^{14}$N \\
H--D geminal & $1.62$ & $4.3$ & CP rate, deuteron; D local field \\
D--D geminal (NHD$_2$, ND$_3$) & $1.62$ & $0.67$ & rotation-driven D relaxation \\
D--D intermolecular (NH$_2$D) & $\simeq3.0$ & $\simeq0.10$ & rotation-driven D relaxation \\
N--N intermolecular & $3.35$ & $0.017$ & negligible \\
\end{tabular}
\end{ruledtabular}
\end{table*}

\section{Cross polarization into one transition of a spin-1}
\label{sec:cp}

\subsection{Single-transition effective field and the match condition}

Apply an rf field at frequency $\omega_{\rm rf}$ near the $(0,-1)$ transition of $X$ (the $(+1,0)$ case is identical) and a spin-locking field at $\wH$ on the protons. In the frame rotating at $\omega_{\rm rf}$ for $X$ and at $\wH$ for $H$, the $X$ Hamiltonian projected onto the $\{|0\rangle,|{-1}\rangle\}$ pair of bond $k$ is, with the single-transition operators $S^{(23)}_\alpha$ \cite{vega1978,vega1977,wokaun1977},
\begin{equation}
\mathcal H_X^{(k)}=\Delta_k\,S_z^{(23)}+\sqrt2\,\woX\,S_x^{(23)},\qquad \Delta_k=\omega_{0,-1}(\theta_k)-\omega_{\rm rf},
\end{equation}
whose effective field has magnitude and tilt
\begin{equation}
\weff^{(k)}=\sqrt{\Delta_k^2+2\woX^2},\qquad \tan\vartheta_k=\frac{\sqrt2\,\woX}{\Delta_k}.
\label{eq:weff}
\end{equation}
The factor $\sqrt2$ is the matrix element $\langle0|S_x|{-1}\rangle$; it is the same factor that enhances the nutation of a single transition of a spin-1 \cite{vega1978} and modifies the spin-locking of quadrupolar nuclei \cite{vega1992,ashbrook2004}. The proton Hamiltonian in its rotating frame is $\woH I_x+\mathcal H_{HH}^{(0)}$, with $\mathcal H_{HH}^{(0)}$ the part of the homonuclear coupling that commutes with $I_x$ (the secular part in the tilted frame, scaled by $-\tfrac12$ \cite{slichter1990,mehring1983}).

In the doubly rotating, doubly tilted frame the heteronuclear term $d_{jk}I_{zj}S_{zk}$ acquires components that connect the proton Zeeman reservoir along $\woH$ with the $X$ fictitious-spin reservoir along $\weff$. Writing the transverse parts, the flip-flop term of the fictitious pair is
\begin{equation}
\mathcal H_{HX}^{\rm ff}=\sum_{jk}\frac{d_{jk}}{2}\,\sin\vartheta_k\,\bigl(I_j^{+'}S_k^{-'(23)}+I_j^{-'}S_k^{+'(23)}\bigr),
\label{eq:ff}
\end{equation}
with primes denoting operators quantized along the respective effective fields, and it conserves energy when
\begin{equation}
\woH=\weff^{(k)}=\sqrt{\Delta_k^2+2\woX^2}
\label{eq:HH}
\end{equation}
to within the proton local field. This is the Hartmann--Hahn condition \cite{hartmann1962} for one transition of a spin-1; on resonance it reads $\woH=\sqrt2\woX$. For a bond at offset $\Delta_k$ the match is met at two values of $\woX$ or, at fixed $\woX$, at $\Delta_k=\pm\sqrt{\woH^2-2\woX^2}$. The width of the match in $\weff$ is of order the proton local field in the tilted frame, $\sim\tfrac12\sqrt{\langle b^2\rangle}$ \cite{demco1975}, several kHz in ammonia, and the width in offset $\Delta$ follows from Eq.~\eqref{eq:weff}.

\subsection{Rotating-frame thermodynamics and the contact map}
\label{sec:contact}

At the match the two reservoirs exchange energy through Eq.~\eqref{eq:ff} and approach a common rotating-frame inverse temperature $\beta_r$ \cite{redfield1955,lurie1964}. The proton spin lock is prepared adiabatically from the laboratory Zeeman order, so $\beta_r\woH=\beta_s\wH$ (adiabatic transfer conserves polarization), i.e.\ the proton polarization along the lock equals $P_H$ and
\begin{equation}
e^{\beta_r\woH}=\frac{1+P_H}{1-P_H}\equiv\rho_H .
\label{eq:rhoH}
\end{equation}
The contacted transition of $X$ is a fictitious spin-$\tfrac12$ with Zeeman splitting $\weff=\woH$ in the tilted frame, so at thermal equilibrium with the lock its two populations, which are the populations of $|0\rangle$ and $|{-1}\rangle$ when the contact is made on resonance ($\vartheta=\pi/2$, the effective field transverse) and then returned adiabatically to the laboratory frame, satisfy
\begin{equation}
\frac{p_0}{p_-}=e^{\beta_r\woH}=\rho_H .
\label{eq:ratio}
\end{equation}
The contact is therefore the map on the three populations of $X$
\begin{equation}
\begin{aligned}
\Phi_{(0,-1)}&:\ (p_+,p_0,p_-)\longmapsto\Bigl(p_+,\ \frac{\rho_H}{1+\rho_H}\,s,\ \frac{1}{1+\rho_H}\,s\Bigr),\\
s&=p_0+p_-.
\end{aligned}
\label{eq:contactmap}
\end{equation}
in the limit that the proton reservoir is unchanged by the contact (Sec.~\ref{sec:heat}), and similarly for $\Phi_{(+1,0)}$ on the upper pair. Inverting the proton lock, which can be done adiabatically or by a $\pi$ pulse on the locked magnetization, replaces $\rho_H$ by $1/\rho_H$ and drives population the other way.

The map is non-unital: it does not preserve the identity and its fixed points are not permutations of the input. This is what lets it pass the bound of the companion paper, which applies to all unital maps. From the EST state at $P_H=0.9$ (Table~\ref{tab:est}, $\rho_H=19$):
\begin{align}
\Phi_{(+1,0)}&:\ (0.490,0.312,0.198)\mapsto\nonumber\\
&\qquad(0.762,0.040,0.198),\nonumber\\
&\qquad\Pzz=+0.88,\ P=0.56;\nonumber\\[0.6em]
\Phi^{\rm inv}_{(+1,0)}\circ\Phi_{(0,-1)}&:\ (0.490,0.312,0.198)\mapsto\nonumber\\
&\qquad(0.049,0.926,0.026),\nonumber\\
&\qquad\Pzz=-1.78,\ P=0.02.
\label{eq:examples}
\end{align}
against the unitary bound $-0.85\le\Pzz\le+0.65$ from the same vector polarization and the saturation ceiling of semi-saturation methods \cite{keller2020,keller2026}. For $^{14}$N a single contact on either transition gives a population ratio of $19$ across it; both transitions contacted in sequence with the same lock sign give $P_N\to0.9$ in the small-reservoir limit.

\subsection{Idempotence and its consequences}
\label{sec:idempotent}

\paragraph*{Proposition.} $\Phi_{(a,b)}\circ\Phi_{(a,b)}=\Phi_{(a,b)}$, and $\Phi_{(a,b)}$ commutes with any map that leaves $p_a+p_b$ fixed.

\paragraph*{Proof.} $\Phi_{(a,b)}$ leaves $s=p_a+p_b$ and the third population unchanged and sets $(p_a,p_b)$ to $s\,(\rho,1)/(1+\rho)$, which depends only on $s$; applying it again returns the same pair. $\square$

Three consequences distinguish the scheme from pulsed inversion \cite{companion}.
\begin{enumerate}
\item \emph{No hit-once condition.} A bond carried through the window twice, or held in it for an arbitrary time, ends in the same state as a bond carried through once. Caps swept at successive orientations may overlap freely, and the set of bonds polarized by a sequence of contacts is the \emph{union} of the caps, not their symmetric difference. Continuous rotation with a continuous contact is allowed.
\item \emph{No parity theorem.} The parity theorem of Ref.~\cite{companion} rests on an adiabatic passage being an involution, so that an even number of crossings is the identity. A bath contact is not an involution, and the net effect of any history of crossings of the window is the projection $\Phi$, whatever the number of crossings.
\item \emph{Robustness to scrambling.} Dipolar flip-flops between $X$ spins, static or rotation-driven, permute single-transition states among bonds and conserve $s$ for the ensemble, so by the proposition a subsequent contact restores the matched bonds regardless of the scrambling in between. The same-branch spin diffusion that converts a sequence of inversions into saturation \cite{companion} is harmless here, and for $^{14}$N it is absent anyway (Table~\ref{tab:couplings}).
\end{enumerate}
What idempotence does not remove is relaxation of the populations \emph{toward a different fixed point}: lattice relaxation and, for deuterons, cross-branch flip-flops that drive the populations back to the EST manifold $p_0^2=p_+p_-$. These compete with the contact rate and are treated in Sec.~\ref{sec:steady}.

\subsection{Heat capacities and proton depletion}
\label{sec:heat}

Equation~\eqref{eq:contactmap} assumes the proton reservoir is not changed by the contact. In the rotating frame the two reservoirs exchange energy quanta of $\woH$, so each flip-flop that moves one $X$ spin from $|{-1}\rangle$ to $|0\rangle$ moves one proton from along the lock to against it. If $N_X^{\rm w}$ bonds are in the window and $N_H$ protons are locked, the proton polarization along the lock falls by
\begin{equation}
\delta P_H=\frac{N_X^{\rm w}}{N_H}\,\delta(p_0-p_-)\,,
\label{eq:depletion}
\end{equation}
per contact, and the common final temperature is found by energy conservation: with $N_H$ spin-$\tfrac12$ at $\rho_H$ and $N_X^{\rm w}$ fictitious spin-$\tfrac12$ initially at ratio $\rho_X$, the final ratio $\rho_f$ solves $N_H\tanh(\tfrac12\ln\rho_f)+N_X^{\rm w}\tanh(\tfrac12\ln\rho_f)=N_H\tanh(\tfrac12\ln\rho_H)+N_X^{\rm w}\tanh(\tfrac12\ln\rho_X)$, i.e.
\begin{equation}
P_f=\frac{N_H P_H+N_X^{\rm w}P_X^{(ab)}}{N_H+N_X^{\rm w}},\qquad \rho_f=\frac{1+P_f}{1-P_f},
\label{eq:Pf}
\end{equation}
where $P_X^{(ab)}=(p_a-p_b)/(p_a+p_b)$ is the initial fictitious polarization of the contacted pair. In NH$_3$ a horn window of width $\delta c$ contains a fraction $2g(\pm1)\,\delta c$ of the nitrogens (Sec.~\ref{sec:geometry}), of order a few percent, and $N_H=3N_N$, so $N_X^{\rm w}/N_H\lesssim10^{-2}$ and the depletion per contact is below $1\%$. Over a full turn the cumulative depletion is $N_X/N_H$ times the transferred polarization, about $0.3$ for nitrogen in NH$_3$, which DNP must restore; the DNP build-up time of minutes sets the pace of the contact train. In NH$_2$D, $N_D/N_H=\tfrac12$ and the same accounting applies with a smaller reservoir ratio. The laboratory-frame heat capacities, $\propto N\gamma^2I(I+1)$, are $215:1$ for H:N in NH$_3$ and $32:1$ for H:D in NH$_2$D, so the lock reservoir dominates throughout.

\subsection{Rotor-driven adiabatic contact}
\label{sec:adiabatic}

BRAIN-CP \cite{harris2012} sweeps $\omega_{\rm rf}$ across the $X$ line during the proton lock; each bond's $\weff$ then passes through $\woH$ and, if the passage is slow compared with the inverse coupling, the $X$ fictitious spin follows its effective field and takes up the proton polarization adiabatically. Rotation does the same thing to a fixed-frequency contact: $\Delta_k(t)=\wX+3\wQ c_k(t)-\omega_{\rm rf}$ sweeps as $c_k(t)$ changes, and $\weff^{(k)}(t)$ crosses $\woH$ when $|\Delta_k|=\sqrt{\woH^2-2\woX^2}$. The condition for the passage to transfer fully is that the time spent within the match width $\delta\weff\simeq\tfrac12\sqrt{\langle b^2\rangle}$ exceeds the transfer time $\sim(d_{\rm eff})^{-1}$, with $d_{\rm eff}=\tfrac12 d\sin\vartheta$ from Eq.~\eqref{eq:ff}:
\begin{equation}
\frac{\delta\weff}{|\dd\weff/\dd t|}\gtrsim\frac{1}{d_{\rm eff}},\qquad
\Bigl|\frac{\dd\weff}{\dd t}\Bigr|=\frac{|\Delta_k|}{\weff}\,3\wQ\,|\dot c_k| .
\label{eq:adiabaticCP}
\end{equation}
This has the form of the Landau--Zener adiabaticity of a two-level crossing with coupling $d_{\rm eff}$ and rate $\dd\weff/\dd t$ \cite{landau1932,zener1932}, and for a single $X$ spin coupled to a proton bath the transfer probability is well approximated by $1-\exp(-2\pi d_{\rm eff}^2/|\dd\weff/\dd t|)$; the many-body proton reservoir makes the transfer irreversible rather than oscillatory \cite{demco1975,mcarthur1969}. Because the process is a bath contact, a passage that is only partly adiabatic transfers part of the polarization and leaves the rest to the next passage; there is no analogue of the mixed state that a partly adiabatic inversion leaves behind.

\section{Reach of a fixed window under rotation}
\label{sec:geometry}

\subsection{Bond trajectories}

Rotating the sample about a laboratory axis $\hat a$ at angle $\beta$ to $B_0$ moves the field direction on a cone of half-angle $\beta$ in the sample frame, and a bond at angle $\alpha_u$ to $\hat a$ sees
\begin{equation}
\cos\theta(t)=\cos\beta\cos\alpha_u+\sin\beta\sin\alpha_u\cos(\Omega t+\phi_u),
\label{eq:traj}
\end{equation}
so $\theta(t)$ oscillates between $\theta_{\min}=|\beta-\alpha_u|$ and $\theta_{\max}=\min(\beta+\alpha_u,\,180\degree-\beta-\alpha_u)$, and $c(t)=3\cos^2\theta(t)-1$ between the corresponding values. Since only $|\cos\theta|$ enters $c$, a bond and its antipode are equivalent and $\alpha_u$ may be taken in $[0,90\degree]$ with the isotropic weight $\sin\alpha_u\,\dd\alpha_u$.

\subsection{Reachable fraction}

\begin{figure}
\includegraphics[width=\columnwidth]{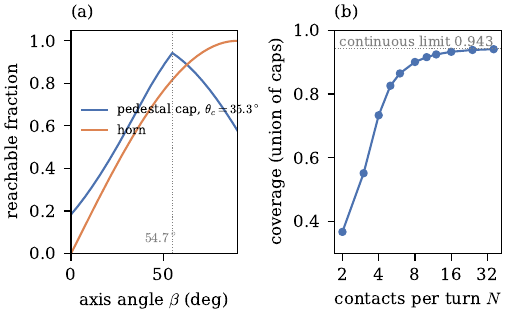}
\caption{(a) Fraction of bonds carried through a horn window ($\sin\beta$) and through the clean pedestal cap (Eq.~\eqref{eq:fcap}) under continuous rotation about an axis at angle $\beta$ to $B_0$; the cap reach peaks at $\beta=54.74\degree$ with $0.943$. (b) Coverage of $N$ pedestal contacts per turn at $\beta=54.74\degree$, the union of $N$ caps of half-angle $\theta_c$ on the cone (Monte Carlo over $2\times10^5$ isotropic bonds), approaching the continuous limit.}
\label{fig:reach}
\end{figure}

A window at reduced frequency $R_w$ on one branch is occupied by bonds at the single angle $\theta_w$ with $3\cos^2\theta_w-1=\pm R_w$. A bond is carried through the window if $\theta_w\in[\theta_{\min},\theta_{\max}]$ or $180\degree-\theta_w$ is. For $\theta_w\le90\degree$ and $\beta\le90\degree$ this is $\alpha_u\in[\,|\theta_w-\beta|,\ \theta_w+\beta\,]$ (the mirrored interval lies above $90\degree$ whenever $\theta_w+\beta\le90\degree$, and coincides with it otherwise), so the reachable fraction is
\begin{equation}
f_w(\beta)=\cos\bigl(\max(0,\theta_w-\beta)\bigr)-\cos\bigl(\min(90\degree,\theta_w+\beta)\bigr).
\label{eq:fw}
\end{equation}
A \emph{band} of windows, such as a contact that is swept over or broad enough to cover the whole clean pedestal $\theta\le\theta_c$, is reached by every bond whose $\theta_{\min}\le\theta_c$, i.e.\ $|\beta-\alpha_u|\le\theta_c$, which gives the cap-reach function
\begin{equation}
f_{\rm cap}(\beta)=\cos\bigl(\max(0,\beta-\theta_c)\bigr)-\cos\bigl(\min(90\degree,\beta+\theta_c)\bigr).
\label{eq:fcap}
\end{equation}
Both are plotted in Fig.~\ref{fig:reach}(a).

\paragraph*{Horn window.} With $\theta_w=90\degree$, Eq.~\eqref{eq:fw} gives $f_{\rm horn}(\beta)=\sin\beta$: every bond except those along the axis is carried through the horn when the axis is perpendicular to the field. At the horn both branches coincide at $|R|=1$ and the window also contains the other branch's bonds at $\theta=35.26\degree$, so a horn contact polarizes both transitions of different bonds; for vector polarization this is what is wanted, and for $^{14}$N it makes $\beta=90\degree$ the natural geometry.

\paragraph*{Pedestal band.} For tensor order the sign of the contact must be definite, which confines the window to one clean pedestal, $\theta\le\theta_c$. Equation~\eqref{eq:fcap} is maximized where $\beta+\theta_c=90\degree$, since for $\beta\ge90\degree-\theta_c$ it falls as $\cos(\beta-\theta_c)$ and for $\beta\le90\degree-\theta_c$ it rises as $2\sin\beta\sin\theta_c$. Hence
\begin{equation}
\beta_{\rm opt}=90\degree-\theta_c=54.74\degree,\quad f_{\rm cap}^{\max}=\sin2\theta_c=\frac{2\sqrt2}{3}=0.943 .
\label{eq:betaopt}
\end{equation}
The magic angle appears for a reason unrelated to the cap packing of Ref.~\cite{companion}: it is the axis tilt at which the cone of field directions is tangent to the equator of the orientation sphere from inside the cap, so the cap sweeps the largest band. The $5.7\%$ of bonds never reached lie within $19.5\degree$ of the rotation axis. With $N$ discrete contacts per turn the coverage is the union of $N$ caps on the cone, Fig.~\ref{fig:reach}(b): $0.73$ for $N=4$ (the tangent cube-diagonal caps), $0.90$ for $N=8$, $0.94$ for $N\ge24$. Overlap costs nothing, by Sec.~\ref{sec:idempotent}.

\section{Rates, the lock, and the steady state}
\label{sec:rates}

\subsection{Dwell time in the window}
\label{sec:dwell}

From Eq.~\eqref{eq:traj}, $\dot c=6\cos\theta\,\dd(\cos\theta)/\dd t$ with $|\dd(\cos\theta)/\dd t|\le\Omega\sin\beta$, so a window of reduced width $\delta c$ is traversed in a time that depends on where it sits.

\paragraph*{Generic position.} Away from the horn, $|\dot c|\le6\Omega\sin\beta$ and
\begin{equation}
\tau_{\rm dwell}\ge\frac{\delta c}{6\,\Omega\sin\beta},\qquad \delta c=\frac{\delta\omega}{3\wQ},
\label{eq:dwelllin}
\end{equation}
linear in $\delta c$. For a deuteron pedestal window of $7$~kHz ($\delta c=0.09$ at $3\wQ/2\pi=75$~kHz) this is $0.019/\Omega$~s, a millisecond at $\Omega=19$~rad\,s$^{-1}$; any slower rotation gives longer dwell.

\paragraph*{Horn.} At $\theta=90\degree$, $\cos\theta=0$ and $\dot c=0$: the horn is a turning point of every trajectory, which is why the powder density diverges there. Near it $c+1\simeq3(\dot x)^2t^2$ with $\dot x=\dd(\cos\theta)/\dd t$, so the time within $\delta c$ of the horn is
\begin{equation}
\tau_{\rm horn}=\frac{2}{|\dot x|}\sqrt{\frac{\delta c}{3}}\ \ge\ \frac{2}{\Omega\sin\beta\sin\alpha_u}\sqrt{\frac{\delta c}{3}},
\label{eq:dwellhorn}
\end{equation}
which scales as $\sqrt{\delta c}$ rather than $\delta c$. For a $30$~kHz $^{14}$N window ($\delta c=0.023$ at $3\wQ/2\pi=1.30$~MHz) this is $0.18/\Omega$~s: $180$~ms at $1$~rad\,s$^{-1}$, far longer than the $\sim0.1$~ms transfer time set by the $8.4$~kHz H--N coupling. The horn is thus the best place to put a fixed window for a rate-limited transfer, independently of its population advantage.

\subsection{The contact train and the proton lock}
\label{sec:lock}

\begin{figure}
\includegraphics[width=\columnwidth]{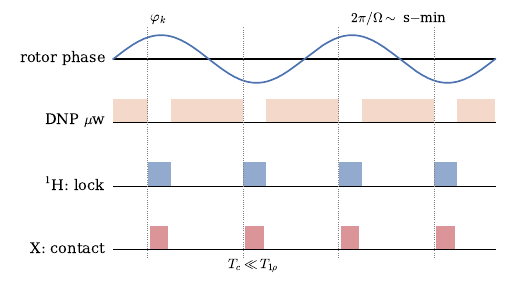}
\caption{Contact train. The sample rotates continuously at a period of seconds to minutes; at rotor phases $\varphi_k$ the microwaves are gated off, the protons are spin-locked, and the $X$ contact field is applied for $T_c\ll T_{1\rho}$; between contacts DNP restores the proton polarization.}
\label{fig:timing}
\end{figure}

The proton spin lock cannot be held for a rotor period. Its decay time $T_{1\rho}$ in a rigid solid is set by slow motions and paramagnetic centres and is unmeasured in irradiated ammonia at $1$~K; in rigid dielectric solids at low temperature it ranges from tens of milliseconds to seconds. Each contact therefore has length $T_c\ll T_{1\rho}$, and the protons lose a fraction $\simeq T_c/T_{1\rho}$ of their polarization per contact in addition to the transfer itself. The sequence is a train of contacts at rotor phases $\varphi_k$ (Fig.~\ref{fig:timing}), each polarizing the bonds in the window at that moment, with DNP microwaves on between contacts to restore the protons. The number of contacts per turn needed for full coverage follows from Fig.~\ref{fig:reach}(b) for a pedestal band and is of order $2\pi/\delta\varphi$ for a narrow window, where $\delta\varphi$ is the rotor phase over which a given bond stays in the window.

The lock amplitude is the hard requirement. The homonuclear coupling in the tilted frame is not removed by the lock, and the locked magnetization survives only if $\woH$ exceeds the proton local field by a wide margin \cite{redfield1955,slichter1990}; in ammonia the geminal H--H coupling alone is $28$~kHz (Table~\ref{tab:couplings}), so $\nu_{1H}\gtrsim50$--$100$~kHz is needed. With $\woH=\sqrt2\woX$ at the match, the $X$ channel then runs at $\nu_{1X}\simeq35$--$70$~kHz. For a polarized-target cell of $\sim15$~cm$^3$ this is a different regime from the milliwatt Q-meter NMR of the targets: scaling from a $20$-turn coil of $2.5$~cm diameter \cite{keller2020}, where $20$--$50$~W gives $\nu_1\simeq2$--$5$~kHz for deuterons and $13$--$33$~kHz for protons, a $100$~kHz proton lock needs of order $0.5$--$1$~kW for the duration of each contact. At $T_c\simeq1$~ms and a few contacts per second the average power is watts, tolerable in a $1$~K helium bath, but the amplifier, coil and feedthroughs are those of a pulsed solid-state NMR probe. A second benefit of a strong proton lock, noted in Ref.~\cite{companion}, is that it decouples the protons from the deuterons during any deuteron rf manipulation; in NH$_2$D the fluctuating $4$~kHz H--D local field would otherwise relax a spin-locked deuteron in a few hundred microseconds.

\subsection{Steady state against relaxation}
\label{sec:steady}

Let the contact train reset the populations of the contacted transition of each reachable bond at a mean rate $k$ (contacts per unit time that find the bond in the window) and let the populations relax toward the EST manifold at a rate $\Gamma$. For the deviation $y=p_a-p_b-(p_a-p_b)^{\rm eq}$ of the contacted pair, with $y_{\rm c}$ its value immediately after a contact, the mean over the cycle is
\begin{equation}
\bar y=y_{\rm c}\,\frac{k}{\Gamma}\bigl(1-e^{-\Gamma/k}\bigr)\ \xrightarrow{\ k\gg\Gamma\ }\ y_{\rm c}\Bigl(1-\frac{\Gamma}{2k}\Bigr),
\label{eq:steady}
\end{equation}
and the ensemble order is the reachable fraction times $\bar y$ plus the EST value for the rest. The relaxation rate $\Gamma$ has three contributions.

\paragraph*{Lattice relaxation.} With microwaves on, DNP itself drives all transitions back toward the common spin temperature at the DNP rate, which is minutes; with microwaves off, $T_1$ at $1$~K is longer still. Both are slow compared with a contact train at seconds.

\paragraph*{Static cross-branch flip-flops.} For $X$ pairs, flip-flops $|0,0\rangle\leftrightarrow|{+1},{-1}\rangle$ with coupling $b'_{kl}/2$ relax the alignment order toward $p_0^2=p_+p_-$ but are quenched in the static sample by the orientation-matching condition $c_k=-c_l$ \cite{companion}. For $^{14}$N the coupling is $17$~Hz and the channel is negligible on any timescale of interest; for deuterons in NH$_2$D the intermolecular coupling is $\simeq0.1$~kHz and the quenched static rate is slow.

\paragraph*{Rotation-driven cross-branch flip-flops.} Rotation sweeps $X$ pairs through $c_k=-c_l$, and the crossing is a Landau--Zener problem with exponent $2\pi V^2/\alpha$, $V=\tfrac12|b'|$ and $\alpha\le2\cdot18\wQ\Omega\sin\beta$ \cite{companion}. For $^{14}$N, $V/2\pi=8$~Hz and $\alpha\simeq4\times10^7\,\Omega$~rad\,s$^{-2}$ ($\wQ/2\pi=434$~kHz), so the exponent is $10^{-4}/\Omega$: diabatic, no transfer, at any practical rate. For deuterons in NH$_2$D with $V/2\pi=50$~Hz and $\wQ/2\pi=25$~kHz, the exponent is $0.25/\Omega$, so rotation at $\Omega\gtrsim1$~rad\,s$^{-1}$ is diabatic and slower rotation is partially adiabatic; each adiabatic crossing between a contacted bond and an uncontacted neighbour costs $\delta\Pzz\simeq3(p_0p_0'-p_+p_-')$ per spin \cite{companion}, of order $0.4$ for the populations of Eq.~\eqref{eq:examples}. The number of such crossings per turn per bond is a few, so for deuterons $\Gamma$ is of order (crossings per turn)$\times$(swap probability)$\times\Omega/2\pi$, and the ratio $\Gamma/k$ in Eq.~\eqref{eq:steady} is small when there are many contacts per turn and the rotation is at or above $1$~rad\,s$^{-1}$. In NHD$_2$ or ND$_3$ the geminal coupling raises $V$ by a factor of seven and the diabatic threshold by fifty, which is the reason the deuteron scheme is specific to NH$_2$D.

\section{Protocols}
\label{sec:protocols}

\subsection{$^{14}$N in NH$_3$}

\begin{enumerate}
\item Rotation axis perpendicular to $B_0$, continuous rotation at $\Omega\simeq0.5$--$2$~rad\,s$^{-1}$.
\item $^{14}$N contact field at a fixed frequency $\wX\pm3\wQ$, one horn ($\pm1.30$~MHz from the nitrogen Larmor frequency, $15.4$~MHz at $5$~T), with $\nu_{1N}\simeq40$--$70$~kHz; a narrowband, high-$Q$ coil tuned once to this frequency.
\item Proton lock $\nu_{1H}=\sqrt2\,\nu_{1N}$, prepared adiabatically from the DNP-polarized Zeeman state, for $T_c\simeq0.5$--$1$~ms $\ll T_{1\rho}$; microwaves gated off during the contact.
\item Contacts at rotor phases spaced by less than the phase width of the horn dwell, Eq.~\eqref{eq:dwellhorn}: with $\tau_{\rm horn}\simeq0.18/\Omega$~s, every $\sim10\degree$ of rotor phase suffices.
\item Alternate the horn sign (or the lock sign) between turns so that both transitions of every nitrogen are contacted; the result is $P_N\to P_H$ in the small-reservoir limit, from $0.14$.
\item Readout by the $^{14}$N NMR area at the horn, calibrated against the EST value with the train off.
\end{enumerate}
The expected gain, from $P_N\simeq0.14$ toward $0.8$--$0.9$, is the largest of any scheme in this paper, and it needs no new material. Nitrogen polarization enters every ammonia-target experiment as a background correction \cite{adeva1998,friedman2019} and a polarized $^{14}$N target has its own physics interest.

\subsection{Deuteron tensor order in NH$_2$D}

\begin{enumerate}
\item Material NH$_2$D, irradiated and DNP-polarized to $P_H\simeq0.9$, deuterons at the EST value $P_D\simeq0.29$ (to be verified, Sec.~\ref{sec:tests}).
\item Rotation axis at $\beta=54.74\degree$, continuous rotation at $\Omega\simeq1$--$3$~rad\,s$^{-1}$ (diabatic for the intermolecular D--D crossings, Sec.~\ref{sec:steady}).
\item Deuteron contact over the clean pedestal of the $(+1,0)$ branch, $-2\le R\le-1$, as a BRAIN-type sweep of $\sim75$~kHz during the lock, so that every bond in the cap passes the match; $\nu_{1D}\simeq35$--$70$~kHz, proton lock $\sqrt2\nu_{1D}$; $T_c\simeq1$--$2$~ms.
\item Contacts at $N\ge8$ rotor phases per turn (coverage $\ge0.90$, Fig.~\ref{fig:reach}(b)); microwaves on between contacts.
\item Single-sign operation gives $\Pzz\to+0.88\times0.94+0.065\times0.06\simeq+0.83$ per Eq.~\eqref{eq:examples} in the ideal limit, with $P\simeq0.55$. For negative tensor order, alternate contacts on the $(0,-1)$ pedestal with the normal lock and on the $(+1,0)$ pedestal with the inverted lock; the ideal limit is $\Pzz\simeq-1.7$ with $P\simeq0$.
\item Readout from the area difference $C(I_+-I_-)$, valid for any population distribution \cite{dulya1997}; the intensity-ratio formulas do not apply off the EST manifold.
\end{enumerate}

\section{Measurements that decide the design}
\label{sec:tests}

\begin{enumerate}
\item \emph{$T_{1\rho}$ of protons} in irradiated NH$_3$ at $1$~K and $5$~T as a function of lock amplitude: a spin-lock decay measurement with the lock prepared from the DNP state. This sets $T_c$, the proton cost per contact, and the lock power.
\item \emph{A single static contact} in NH$_3$: $^{14}$N contact at one horn with the sample at rest, readout of the $^{14}$N line. The horn window contains a known fraction of the nitrogens, and their population ratio should jump from $1.24$ to $\simeq19$; this tests the match, the $\sqrt2$ condition and the transfer time before any rotation.
\item \emph{EST in NH$_2$D}: DNP-polarize and measure $P_H$ and $P_D$ together; Eq.~\eqref{eq:est} predicts $P_D=0.29$ at $P_H=0.90$. A shortfall indicates poor thermal contact of the deuterons, as seen in some BDPA-doped materials \cite{deboer2015}, and would reduce the contact-free baseline but not the contact result.
\item \emph{A single static pedestal contact} in NH$_2$D: the cap holds $18\%$ of the deuterons, so one contact on $(+1,0)$ should raise $\Pzz$ from $0.065$ to $0.18\times0.88+0.82\times0.065=0.21$ and $P$ from $0.29$ to $0.34$; a clean test of Eq.~\eqref{eq:contactmap} and of the readout.
\item \emph{Rotation-driven deuteron relaxation} in NH$_2$D: contact one cap at rest, rotate one turn at a set $\Omega$ without contacts, read the survival. This maps the diabatic threshold of Sec.~\ref{sec:steady}.
\end{enumerate}

\section{The cubic limit and material requirements}
\label{sec:materials2}

The scheme has been presented for a quadrupolar spin-1 whose two transitions can be addressed separately. This section treats the opposite limit, $\wQ\to0$, which is realized at cubic nitrogen sites, and derives from the two limits the conditions a material must meet.

\subsection{Contact in the limit $\wQ\to0$}
\label{sec:cubic}

At a site of cubic symmetry the electric field gradient vanishes and Eq.~\eqref{eq:H} reduces, for $X$, to $-\wX S_z$. An rf field is then a true SU(2) rotation of the spin-1, both single-quantum transitions are degenerate, and the rotating-frame Hamiltonian is $\woX S_x$ with equal splittings $\woX$ between adjacent eigenstates of $S_x$. The Hartmann--Hahn condition is $\woH=\woX$, without the $\sqrt2$ of Eq.~\eqref{eq:HH}, and the flip-flop term of Eq.~\eqref{eq:ff} connects adjacent $S_x$ levels with the spin-1 matrix elements. At equilibrium with the lock both adjacent-level ratios equal $e^{\beta_r\woH}=\rho_H$, and after adiabatic return to the laboratory frame the populations are
\begin{equation}
\Phi_{\rm full}:\ (p_+,p_0,p_-)\longmapsto\frac{(\rho_H,\,1,\,1/\rho_H)}{\rho_H+1+1/\rho_H},
\label{eq:fullcontact}
\end{equation}
independent of the input. This is a spin-temperature state with $a_X=\ln\rho_H=a_H$: the contact sets the $X$ Zeeman temperature equal to the \emph{proton} Zeeman temperature, replacing the EST scaling $a_X=(\gamma_X/\gamma_H)a_H$ by $a_X=a_H$. From Eq.~\eqref{eq:est},
\begin{equation}
P_X=\frac{\rho_H-\rho_H^{-1}}{\rho_H+1+\rho_H^{-1}},\qquad
\Pzz=\frac{\rho_H+\rho_H^{-1}-2}{\rho_H+1+\rho_H^{-1}},
\label{eq:cubicPzz}
\end{equation}
which at $P_H=0.90$ ($\rho_H=19$) gives $P_X=0.945$ and $\Pzz=+0.850$, against the EST values $0.141$ and $0.015$ for $^{14}$N. The vector polarization exceeds $P_H$ because a spin-1 at a given adjacent-level ratio is more polarized than a spin-$\tfrac12$ (the Brillouin function), and the tensor polarization is the full spin-temperature value $2-\sqrt{4-3P_X^2}$ at that $P_X$. Table~\ref{tab:cubic} gives the dependence on $P_H$.

\begin{table}
\caption{Spin-1 vector and tensor polarization after a full-spin contact in the cubic limit, Eq.~\eqref{eq:cubicPzz}, versus proton polarization.}
\label{tab:cubic}
\begin{ruledtabular}
\begin{tabular}{lccccc}
$P_H$ & $0.50$ & $0.70$ & $0.80$ & $0.90$ & $0.95$ \\
\hline
$P_X$ & $0.615$ & $0.802$ & $0.879$ & $0.945$ & $0.974$ \\
$\Pzz$ & $0.308$ & $0.562$ & $0.703$ & $0.850$ & $0.925$ \\
\end{tabular}
\end{ruledtabular}
\end{table}

\paragraph*{What the cubic limit cannot do.} Every operation available in this limit, the full contact with either lock sign ($\rho_H\to1/\rho_H$ gives $P_X\to-P_X$ and leaves $\Pzz$ unchanged), SU(2) rotations, and lattice relaxation, maps spin-temperature states to spin-temperature states, since each preserves the equality of the two adjacent-level ratios. The accessible set is therefore the one-parameter manifold $p_0^2=p_+p_-$, on which
\begin{equation}
0\le\Pzz=2-\sqrt{4-3P^2}\le\frac{\rho_H+\rho_H^{-1}-2}{\rho_H+1+\rho_H^{-1}} .
\label{eq:cubicbound}
\end{equation}
Tensor order is always positive, is tied to the vector polarization, and cannot exceed the value reached at the proton temperature. Negative tensor order, tensor order without vector order, and the approach to $\Pzz=-2$ of Eq.~\eqref{eq:examples} all require the two transitions to be contacted with \emph{different} lock signs or at different times, which requires them to be resolved: $3\wQ$ must exceed the contact window. This is the precise content of the non-cubic condition for tensor enhancement: a quadrupole splitting is not needed to create large positive alignment from a proton bath, but it is needed to leave the spin-temperature manifold.

\paragraph*{Ammonium chloride.} NH$_4$Cl is the standard $^{14}$N chemical-shift reference because its nitrogen occupies a site of $T_d$ symmetry in both the disordered (CsCl-type) and the ordered low-temperature phase, so the static EFG vanishes and the $^{14}$N line is narrow \cite{jakobsen2006}; calculations of the NQCC of NH$_4$Cl \cite{halkier1999} concern the sensitivity of the local EFG to hydrogen-bond geometry, not a static splitting in the crystal. It is therefore the cleanest possible test of Eq.~\eqref{eq:fullcontact}: no lineshape, no orientation dependence, no rotation, a single matched contact, and a predicted $\Pzz=+0.85$ from $P_H=0.9$ with four protons per nitrogen at $1.03$~\AA. What it lacks is a route to $P_H=0.9$: ammonium chloride is not a DNP target material, and whether irradiation-induced or dopant paramagnetic centres in it support DNP at $1$~K is untested.

\subsection{Requirements for selective (off-manifold) tensor order}
\label{sec:requirements}

Combining the two limits, a material for the selective scheme of Secs.~\ref{sec:cp}--\ref{sec:protocols} must satisfy four conditions.

\paragraph*{(i) Resolvable pedestal.} The clean pedestal has width $(2-1)\times3\wQ=3C_Q/8$ in frequency for $\eta=0$, and the contact addresses a window of width $\delta\omega\simeq\max(\sqrt2\woX,\ \delta\weff)$, tens of kHz. Selectivity requires
\begin{equation}
\frac{3C_Q}{8}\gg\delta\omega\quad\Longrightarrow\quad C_Q\gtrsim\frac{8}{3}\,\kappa\,\delta\omega ,
\label{eq:CQmin}
\end{equation}
with $\kappa\simeq5$ for the window to sit well inside the pedestal: $C_Q\gtrsim0.4$~MHz for a $30$~kHz window, $0.13$~MHz for a $10$~kHz window. Below this the window overlaps the horn of the other branch and the contact acquires contributions of both signs. Ammonia ($3.47$~MHz) satisfies the condition by an order of magnitude; the amino-acid range $\simeq1.2$~MHz satisfies it; ammonium salts span $C_Q\simeq0.02$--$1$~MHz \cite{jakobsen2006} and must be selected.

\paragraph*{(ii) Small asymmetry.} For $\eta\ne0$ the first-order transitions are at $\wX\mp3\wQ(3\cos^2\theta-1+\eta\sin^2\theta\cos2\phi)$ \cite{cohen1957}, so the lower edge of each branch moves from $R=\mp1$ to $R=\mp(1+\eta)$ while the pedestal edge stays at $R=\pm2$. The clean region shrinks to $1+\eta<|R|\le2$, the cap angle to
\begin{equation}
\cos^2\theta_c(\eta)=\frac{2+\eta}{3},
\label{eq:thetaeta}
\end{equation}
and the cap fraction $1-\cos\theta_c$ from $0.184$ at $\eta=0$ to $0.124$ at $\eta=0.3$, $0.087$ at $\eta=0.5$, and zero at $\eta=1$, where the two branches overlap completely and no selective contact exists. The optimal axis angle of Eq.~\eqref{eq:betaopt} becomes $90\degree-\theta_c(\eta)$ and the reach $\sin2\theta_c(\eta)$. Sites of $C_3$ or higher axial symmetry, N--H in ammonia and the pseudo-axial N sites of some ammonium salts, have $\eta\simeq0$; low-symmetry ammonium sites can have any $\eta$ up to $1$ \cite{jakobsen2006}, so $\eta$ is a selection criterion as important as $C_Q$.

\paragraph*{(iii) A polarizable proton bath.} The tensor order reachable from an unpolarized spin-1 by one selective contact is
\begin{equation}
\Pzz=\frac{\rho_H-1}{\rho_H+1}=P_H ,
\label{eq:PzzeqPH}
\end{equation}
and by a full contact Eq.~\eqref{eq:cubicPzz}, so the absolute tensor polarization is bounded by the proton polarization available. At the $P_H\simeq0.01$--$0.05$ of magic-angle-spinning DNP at $\sim100$~K the signal enhancement over thermal $^{14}$N is of order $10^4$--$10^6$, but the absolute alignment is a few percent. Large tensor order in the polarized-target sense requires $P_H\gtrsim0.8$, i.e.\ DNP at $\lesssim1$~K and $\gtrsim5$~T with a radical concentration of order $10^{19}$~cm$^{-3}$ \cite{crabb1997,goertz2002}, which has been achieved in ammonia by irradiation \cite{meyer2004} and in alcohols by doping \cite{crabb1997}. For a crystalline ammonium salt the natural route is irradiation, which is known from EPR to create paramagnetic centres in several ammonium salts, but DNP from those centres has not been demonstrated.

\paragraph*{(iv) Proton lock and reservoir.} The lock must exceed the proton local field, $28$~kHz for geminal N--H protons in ammonia and comparable in the NH$_4^+$ ion ($1.03$~\AA, six H--H pairs at $1.68$~\AA), and the proton-to-nitrogen reservoir ratio sets the depletion per contact, Eq.~\eqref{eq:depletion}: $N_H/N_X=3$ for ammonia and $4$ for ammonium salts, both favourable.

\begin{table*}
\caption{Candidate materials for $^{14}$N (and deuteron) order pumped from a proton bath. ``Selective'' means Eq.~\eqref{eq:CQmin} is satisfied for a $30$~kHz window. $C_Q$ values for the ammonium salts are the range reported by $^{14}$N MAS NMR \cite{jakobsen2006}; individual values and $\eta$ must be taken from that work or measured.}
\label{tab:materials}
\begin{ruledtabular}
\begin{tabular}{p{2.9cm}cccc p{5.2cm}}
Material & N site & $C_Q$ & $\eta$ & $N_H/N_X$ & Status / role \\
\hline
NH$_3$ (irradiated) & $C_{3v}$ & $3.47$~MHz \cite{lehrer1965} & $0$ & $3$ & established target; selective; horn or pedestal; Sec.~\ref{sec:protocols} \\
NH$_2$D (irradiated) & $C_s$ & $\simeq3.5$~MHz (N); $\sim0.2$~MHz (D) & small & $2$ (D) & deuteron tensor order; DNP untested \\
NH$_4$Cl & $T_d$ & $\simeq0$ & -- & $4$ & cubic limit; full contact only, $\Pzz\le+0.85$; no rotation needed; DNP untested \\
NH$_4$H$_2$PO$_4$, (NH$_4$)$_2$HPO$_4$, (NH$_4$)$_2$C$_2$O$_4\cdot$H$_2$O, CH$_3$NH$_3$Cl & low & $0.02$--$1$~MHz \cite{jakobsen2006} & $0$--$1$ & $4$ (3) & selective if $C_Q\gtrsim0.4$~MHz and $\eta\lesssim0.3$; irradiation candidates; DNP untested \\
Glycine, $\alpha$-amino acids & low & $\simeq1.2$~MHz & $\lesssim0.5$ & $3$ & selective; wideline $^{14}$N methods established \cite{odell2009}; DNP at $1$~K untested \\
\end{tabular}
\end{ruledtabular}
\end{table*}

Table~\ref{tab:materials} collects the candidates. Two experimental sequences follow from it. For a demonstration of bath-contact alignment, NH$_4$Cl or any cubic-site salt is the simplest system, provided a proton polarization can be established in it; the measurement is a single contact and a $^{14}$N line integral, with Eq.~\eqref{eq:cubicPzz} as the prediction. For the selective scheme, with its negative and vector-free tensor order, ammonia is the only material in which all four conditions are already known to be met, and the non-cubic ammonium salts with $C_Q$ above $\simeq0.4$~MHz and small $\eta$ are the candidates for extending it to a crystalline solid once their DNP behaviour under irradiation is known. Wideline $^{14}$N acquisition in such salts and in amino acids by WURST and BRAIN-type sequences \cite{odell2009,harris2012} already provides the lineshapes, $C_Q$ and $\eta$ that the geometry of Sec.~\ref{sec:geometry} needs.

\section{Conclusion}
\label{sec:conclusion}

Cross polarization from a highly polarized proton reservoir into one transition of a quadrupolar spin-1 is a bath contact, and a bath contact is an idempotent map on the populations of the contacted transition. That single algebraic fact separates the present scheme from every pulsed-inversion scheme: contacts accumulate, caps may overlap, continuous rotation with a continuous or repeated contact is allowed, dipolar scrambling between contacts is harmless, and the result is not bounded by the spectrum of the initial state, so that tensor polarizations of $+0.88$ or $-1.8$ are reachable from a deuteron ensemble at $P=0.29$, and $^{14}$N can be taken from its equal-spin-temperature value of $0.14$ toward the proton polarization. Slow rotation of the sample carries every crystallite of a MHz-wide line through a single fixed-frequency window, with the reachable fraction given in closed form by Eq.~\eqref{eq:fw}: all bonds for a horn window and a perpendicular axis, $0.943$ for a pedestal band and an axis at the magic angle. The horn is a turning point of each bond's frequency trajectory, so the dwell there is long and favours slow rotation. In the cubic limit the same contact gives the full spin-temperature alignment at the proton temperature, $\Pzz=+0.85$ from $P_H=0.9$, with no rotation at all, so a cubic-site salt such as NH$_4$Cl is the simplest demonstration of bath-contact alignment; what a resolvable quadrupole splitting adds is the freedom to leave the spin-temperature manifold, and the conditions for that, $3e^2qQ/8h$ well above the contact window, small asymmetry, and a proton bath polarizable to $\gtrsim0.8$, select ammonia now and the non-cubic ammonium salts once their DNP behaviour is known. What limits the scheme is not the spin-1 physics but the proton spin lock, which in ammonia must exceed a $28$~kHz local field for a time short compared with a $T_{1\rho}$ that has not been measured at $1$~K; the measurements of Sec.~\ref{sec:tests} are ordered so that this is settled first, with a single static $^{14}$N contact as the decisive early test.

\appendix

\section{Spin-1 populations at a spin temperature}
\label{app:est}

With $p_m=e^{ma}/Z$, $Z=e^{a}+1+e^{-a}=1+2\cosh a$, the vector and tensor polarizations are $P=(e^a-e^{-a})/Z=2\sinh a/(1+2\cosh a)$ and $\Pzz=1-3p_0=(2\cosh a-2)/(1+2\cosh a)$, which eliminate $a$ through $\Pzz=2-\sqrt{4-3P^2}$. The population ratio across either single-quantum transition is $e^{a}$, and the fictitious polarization of the pair is $\tanh(a/2)$. For protons, $P_H=\tanh(a_H/2)$, so $e^{a_H}=(1+P_H)/(1-P_H)=\rho_H$, which is Eq.~\eqref{eq:rhoH} with $\beta_r\woH=\beta_s\wH=a_H$.

\section{Single-transition operators and the $\sqrt2$ factor}
\label{app:sto}

For spin-1 in the basis $(|{+1}\rangle,|0\rangle,|{-1}\rangle)$, $S_x=\tfrac{1}{\sqrt2}\begin{pmatrix}0&1&0\\1&0&1\\0&1&0\end{pmatrix}$. The single-transition operators of the pair $(r,s)$ are $S_x^{(rs)}=\tfrac12(|r\rangle\langle s|+|s\rangle\langle r|)$, $S_y^{(rs)}=\tfrac{1}{2i}(|r\rangle\langle s|-|s\rangle\langle r|)$, $S_z^{(rs)}=\tfrac12(|r\rangle\langle r|-|s\rangle\langle s|)$, which obey spin-$\tfrac12$ commutation relations within the pair \cite{vega1978,wokaun1977}. Then $S_x=\sqrt2\,(S_x^{(12)}+S_x^{(23)})$, so an rf field $\woX S_x$ acts on each pair as a fictitious spin-$\tfrac12$ with Rabi frequency $\sqrt2\woX$, and the quadrupole term $\wQ c(3S_z^2-2)$ is diagonal with $\langle{+1}|\cdot|{+1}\rangle=\langle{-1}|\cdot|{-1}\rangle=\wQ c$, $\langle0|\cdot|0\rangle=-2\wQ c$, giving the offsets $\mp3\wQ c$ of the two transitions. The heteronuclear coupling $d\,I_zS_z$ restricted to the pair is $d\,I_z(S_z^{(23)}-\tfrac12)$ for $(0,-1)$, whose $I_zS_z^{(23)}$ part supplies the flip-flop term of Eq.~\eqref{eq:ff} after tilting both spins to their effective fields; the constant shifts the pair's centre and drops out.

\section{The contact map as a projection}
\label{app:projection}

On the simplex of population vectors $(p_+,p_0,p_-)$ the map $\Phi_{(0,-1)}$ of Eq.~\eqref{eq:contactmap} is affine on each fibre $\{p_0+p_-=s,\ p_+=1-s\}$ and constant on it, hence a projection onto the curve $\{(1-s,\ \rho s/(1+\rho),\ s/(1+\rho))\}$. Its image is one-dimensional, so it is not injective, not invertible and not unital (the uniform vector $(\tfrac13,\tfrac13,\tfrac13)$ maps to $(\tfrac13,\tfrac23\rho/(1+\rho),\tfrac23/(1+\rho))$). The Schur--Horn bound of Ref.~\cite{companion}, which states that the output populations of a unital map are majorized by the input spectrum, therefore does not apply; indeed for $\rho=19$ the output $(0.762,0.040,0.198)$ of Eq.~\eqref{eq:examples} is not majorized by $(0.490,0.312,0.198)$.

The composition $\Phi^{\rm inv}_{(+1,0)}\circ\Phi_{(0,-1)}$ is also a projection, onto the single point $\bigl(1/(1+\rho)^2\cdot(\ldots)\bigr)$ that solves $p_0/p_-=\rho$ and $p_0/p_+=\rho$ simultaneously, i.e.\ $(p_+,p_0,p_-)=(1,\rho,1)/(2+\rho)$, with $\Pzz=1-3\rho/(2+\rho)=(2-2\rho)/(2+\rho)$, which is $-1.71$ at $\rho=19$ and tends to $-2$ as $\rho\to\infty$. The value $-1.78$ in Eq.~\eqref{eq:examples} is the single application of the two maps starting from the EST state; repeated alternation converges to $-1.71$, the fixed point, since the second map disturbs the fibre of the first.

\section{Reach: derivation of Eq.~\eqref{eq:fw}}
\label{app:reach}

The extremes of $\theta(t)$ in Eq.~\eqref{eq:traj} occur at $\cos(\Omega t+\phi_u)=\pm1$ and are $\cos\theta_\pm=\cos(\beta\mp\alpha_u)$, so $\theta$ ranges over $[\,|\beta-\alpha_u|,\ \beta+\alpha_u\,]$ for $\beta+\alpha_u\le180\degree$. The window at $\theta_w$ (or its mirror $180\degree-\theta_w$, equivalent under $\theta\to180\degree-\theta$) is visited iff $|\beta-\alpha_u|\le\theta_w\le\beta+\alpha_u$ or $|\beta-\alpha_u|\le180\degree-\theta_w\le\beta+\alpha_u$. The first condition is $\alpha_u\in[\,|\theta_w-\beta|,\ \theta_w+\beta\,]$; the second is $\alpha_u\in[\,|180\degree-\theta_w-\beta|,\ 180\degree-\theta_w+\beta\,]$, whose lower end is $\ge90\degree$ when $\theta_w+\beta\le90\degree$ and otherwise is $\le90\degree$ but contained in the reflection of the first interval about $90\degree$. Restricting $\alpha_u$ to $[0,90\degree]$ with weight $\sin\alpha_u$ and integrating gives Eq.~\eqref{eq:fw}. For a cap $\theta\le\theta_c$ the condition is $\theta_{\min}=|\beta-\alpha_u|\le\theta_c$, which gives Eq.~\eqref{eq:fcap}; the maximum over $\beta$ of $\cos(\beta-\theta_c)-\cos(\min(90\degree,\beta+\theta_c))$ occurs at $\beta=90\degree-\theta_c$ where both branches of the function meet with value $\cos(90\degree-2\theta_c)=\sin2\theta_c$.

\bibliographystyle{apsrev4-2}
\bibliography{refs}

\end{document}